\documentclass[namedreferences]{SolarPhysics}
\usepackage[optionalrh]{sp-addons} 
\usepackage{epsfig}          
\usepackage{graphicx}        
\usepackage{courier}         
\usepackage{amssymb}        
\usepackage{color}           
\usepackage{url}             
\newcommand{\etal}{{\it et al.}}
\newcommand{\be}{\begin{equation}}
\newcommand{\ee}{\end{equation}}
\newcommand{\beq}{\begin{eqnarray}}
\newcommand{\eeq}{\end{eqnarray}}

\begin{document}
\begin{article}
\begin{opening}

\title{Velocity-Space Proton Diffusion in the Solar Wind Turbulence}

\author{Y.~Voitenko$^{1}$ \sep
V.~Pierrard$^{1,2}$
   }

\runningauthor{Voitenko and Pierrard}
\runningtitle{Velocity-Space Proton Diffusion in the Solar Wind Turbulence}

\institute{$^1$ Solar-Terrestrial Centre of Excellence, Space Physics Division,
Belgian Institute for Space Aeronomy, Ringlaan-3-Avenue Circulaire, B-1180 Brussels, Belgium,
email: \url{yuriy.voitenko@oma.be}
\\
$^{2}$  Georges Lema\^\i tre Centre for Earth and Climate Research (TECLIM),
Universit\'e Catholique de Louvain, Place
Louis Pasteur 3, bte L4.03.08, 1348 Louvain-la-Neuve, Belgium\\}

\date{Published in ArXiv 3 April 2013}

\begin{abstract}
We study a velocity-space quasilinear diffusion of the solar wind protons
driven by oblique Alfv\'{e}n turbulence at proton kinetic scales.
Turbulent fluctuation at these scales possess properties of kinetic Alfv\'{e}n
waves (KAWs) that are efficient in Cherenkov resonant interactions. The
proton diffusion proceeds via Cherenkov kicks and forms a quasilinear plateau
- nonthermal proton tail in the velocity distribution function (VDF).
The tails extend in velocity space along the mean magnetic field
from 1 to (1.5--3) $V_{\rm A}$, depending on the spectral break position,
turbulence amplitude at the spectral break, and spectral slope
after the break. The most favorable conditions for the tail generation
occur in the regions where the proton thermal and Alfv\'{e}n velocities are
about the same, $V_{{\rm Tp}}/V_{\rm A}\approx 1$. The estimated formation times are
within 1--2 h for typical tails at 1 AU, which is much shorter than the
solar wind expansion time. Our results suggest that the nonthermal proton
tails, observed in-situ at all heliocentric distances $>0.3$ AU, are formed
in the solar wind locally by the KAW turbulence.
We also suggest that the bump-on-tail features - proton beams,
often seen in the proton VDFs, can be formed at a later evolution
stage of the nonthermal tails by the time-of-flight effects.
\end{abstract}


\end{opening}



\section{Introduction}

Low collisionality of the solar wind (SW) and persistent activity of waves
and turbulence make kinetic wave-particle interactions unavoidable in the SW
modeling. The need for such kinetics has been emphasized in view of numerous
non-thermal features observed in velocity distribution functions (VDFs) of
the solar wind particles (see \inlinecite{Marsch2006} and
references therein). In particular, nonthermal tails, beams, and temperature
anisotropies are routinely observed in situ by satellites.

Recent in-situ measurements have revealed that the SW turbulence at proton
kinetic scales is dominated by Alfv\'{e}nic waves (AWs) rather than the fast
mode or whistler waves (\opencite{He2011}, \citeyear{He2012}; \opencite{Podesta2011}).
Furthermore, among Alfv\'{e}n waves, the most power, about 80\%, was found in oblique
quasi-perpendicular AWs, and the rest 20\% in the quasi-parallel
ion-cyclotron Alfv\'{e}n waves (ICAW) \cite{He2011}.

Properties of
quasi-parallel ICAWs are understood much better than that of the oblique AWs
at proton kinetic scales. ICAWs experience strong ion-cyclotron resonances
and are considered as a main source for the perpendicular heating of the
protons and heavier ions in the solar corona and solar wind (see reviews
by \inlinecite{Hollweg2002}, \inlinecite{Marsch2006} and references therein).

Properties of the dominant quasi-perpendicular AW fraction and its influence
on particles VDFs are still unclear. Depending on the wavevector anisotropy $%
k_{\perp }/k_{\parallel }$ (perpendicular/parallel with respect to the mean
magnetic field $B_{0}$), less oblique AWs at proton kinetic scales are
similar to ICAWs, while more oblique AWs are kinetic Alfv\'{e}n waves
(KAWs). At intermediate $k_{\perp }/k_{\parallel }\approx V_{\rm A}/V_{{\rm Tp}}$,
AWs possess mixed
properties of both ICAWs and KAWs and can be named ion-cyclotron KAWs -
ICKAWs \cite{Voitenko2003}. Since ICKAWs experience both the
ion-cyclotron resonance (typical for ICAWs) and the Cherenkov resonance
(typical for KAWs), they have more dissipation and generation channels and
can play an important role in the energy exchange between parallel and
perpendicular degrees of freedom in the solar wind.

The wavevector anisotropy at proton kinetic scales has been measured by
\inlinecite{Sahraoui2010}. For perpendicular wave numbers up to $k_{\perp }\rho
_{\rm p}\simeq 2$, the wave/proton-cyclotron frequency ratio appeared to be
still low, $\omega /\Omega _{\rm p}\simeq 0.1$, whereas the anisotropy is high,
about $k_{\perp }/k_{\parallel }\simeq 10$, which is typical for classic
KAWs. The dominant role of KAWs in the oblique AW fraction has been
supported further by two independent tests performed by \inlinecite{Salem2012}, who
measured simultaneously two polarization ratios: compressional magnetic/total magnetic
and total electric/total magnetic field perturbations.

The interest in KAWs and related issues has risen recently in the context of
turbulence dissipation in the solar wind and consequent heating of plasma
species (\inlinecite{Schekochihin2009}; \inlinecite{Howes2011}; and references therein).
\inlinecite{Goldreich1995} suggested that the MHD AW turbulence cascades
along a critical balance path, which result in preferential generation of
high perpendicular wavenumbers, and hence KAWs \cite{Schekochihin2009}.
An indirect confirmation of the critical
balance was received recently by \inlinecite{Gogoberidze2012},
who proved that the critical balance leads to generation of the residual energy
observed in the solar wind turbulence.

A solar wind model elaborating Landau damping of KAWs has predicted the
electron and proton heating rates \cite{Howes2011}, which agree well with
empirical estimations by \inlinecite{Cranmer2009} at $\gtrsim 0.8$ AU. At
smaller radial distances the proton heating by Landau damping appeared to be
insufficient and Howes (2011) suggested the proton cyclotron heating
operating there. Because of the shorter inertial range, the MHD turbulence
at shorter radial distances ends up with less anisotropic fluctuations
at proton kinetic scales, which can produce the required proton heating
via cyclotron resonant interaction.

A widely used assumption of the Maxwellian Landau damping can be violated
in the solar wind by local deformations of VDFs of particles in resonant
velocity ranges. \inlinecite{Borovsky2011} have demonstrated that the electron
Landau damping is still strong (although not Maxwellian any more)
because the relatively high collisionality of the electron is capable of keeping
their VDFs not far from Maxwellian. On the contrary, the less collisional protons
experience much stronger departures from Maxwellian VDFs, which can significantly
modify the proton Landau damping, or even cancel it.

\inlinecite{Chandran2010} proposed another promising heating mechanism for
the protons, stochastic acceleration by KAWs across $B_{0}$.
This mechanism is not related to the kinetic wave-particle resonances,
and is therefore not so sensitive to the local deformations of the proton VDFs.

If the turbulence at proton kinetic scales consists of the ion-cyclotron and
KAW fractions, as is suggested by recent observations, both these
fractions should produce their own observed signatures - specific features
in the proton VDFs. The most typical observed features of the proton VDFs
are \cite{Marsch1991}: (i) nonthermal tails or even secondary peaks along the
magnetic field direction in all solar wind types (fast, slow, and
intermediate), and (ii) total temperature anisotropies $T_{{\rm p}\perp
}>T_{{\rm p}\parallel }$\ in the fast solar wind (with anisotropic proton cores),
and $T_{{\rm p}\parallel }>T_{{\rm p}\perp }$\ in the slow (with isotropic cores) and
intermediate (with both isotropic and anisotropic cores) solar winds. The
parallel proton tails and beams are important contributors to the energy
balance between parallel and perpendicular degrees of freedom. In the fast
solar wind they reduce the total temperature anisotropy to $T_{{\rm p}\perp
}/T_{{\rm p}\parallel }\gtrsim 1$\ (and at times even reverse to $T_{{\rm p}\parallel
}>T_{{\rm p}\perp }$) despite the strongly\ anisotropic cores of the proton VDFs, $%
T_{{\rm p}\perp }^{\mathrm{core}}/T_{{\rm p}\parallel }^{\mathrm{core}}=$\ $2 - 3$
(see Figure 8.4 by Marsch, 1991). In the intermediate and slow solar wind they
dominate the energy balance making $T_{{\rm p}\parallel }/T_{{\rm p}\perp }>1$. Among
them, the nonthermal tails prevail \cite{Marsch1982}. The
secondary peaks (bumps-on-tails) are less frequent and
develop on the tail background mainly in the fast solar wind. Therefore the
bump-on-tail seems to be a later evolution stage of the proton VDF as
compared to the tail. At the same time, persistent nonthermal tails without
bumps are observed at all distances, which suggest that the bumps are not necessary
developing on the tails within the solar wind expansion time-scale; their
development is therefore somehow constrained or require more time.

The anisotropic cores of the proton VDFs with $T_{{\rm p}\perp }^{\mathrm{core}%
}>T_{{\rm p}\parallel }^{\mathrm{core}}$\ have been associated with the
ion-cyclotron heating by ion-cyclotron resonant waves, and ICAWs interaction
with solar wind protons and heavier ions have been studied extensively
(\inlinecite{Hollweg2002}; \inlinecite{Marsch2006}; and references therein). It was in
particular shown that the ion-cyclotron pitch-angle diffusion can be
responsible for the shaping and perpendicular heating of the core proton
VDFs \cite{Marsch2001}, for the cross-field heating of heavier ions
\cite{Galinsky2012}, and also for the pitch-angle scattering of
the nonthermal proton tails \cite{Marsch2011}.

Wave mechanisms producing nonthermal proton tails and bumps-on-tails (beams)
in the solar wind received much less attention.
The first one, proposed by
\inlinecite{Tu2002}, relied upon the proton cyclotron-resonant diffusion driven
by a specific ion-cyclotron mode supported by helium ions.
Another
mechanism, nonlinear proton trapping in the parallel KAW potentials, has
been suggested by \inlinecite{Voitenko2006}. Later on, analytical
\cite{PiVo2010} and numerical \cite{Li2010,Osmane2010}
studies have confirmed that a KAW with reasonable amplitude can
produce observed proton beams in the solar wind conditions.

Proton trapping and beam generation can also be produced by electrostatic
waves carrying parallel electric fields. Recent simulations of the nonlinear
decay of circularly polarized Alfv\'{e}n waves have demonstrated that the
product electrostatic waves can trap and accelerate protons along $B_{0}$,
thereby creating a beam \cite{Araneda2008,Matteini2011,Valentini2011}.
Ion-acoustic waves \cite{Araneda2008,Matteini2011} and "ion-bulk" waves
\cite{Valentini2011} have been
discussed as mediators transferring energy from AWs to plasma particles.
Alternatively, \inlinecite{Rudakov2012} used a quasilinear approach to study
the proton diffusion driven by a strongly dispersive KAW spectrum. This
process can produce a quasilinear plateau in the form of nonthermal tail in
the proton VDF.

Earlier kinetic models of the SW electrons, accounting for the influence of
external macroscopic forces and Coulomb collisions
\cite{Pierrard1999,Lie-Svendsen2000,Vocks2009},
have been recently improved by the inclusion of terms due to
whistler turbulence \cite{PiLa2011}. It was concluded that the whistler
turbulence, if exists in the solar wind, could lead to the velocity-space
diffusion of electrons and form nonthermal tails in the electron VDF.
Another source for the velocity-space electron diffusion has been studied by
\inlinecite{Rudakov2011}, who assumed a spectrum of kinetic Alfv\'{e}n waves (KAWs)
instead of whistlers. Moreover, \inlinecite{Rudakov2011} argued that the nonlinear
KAW scattering off electrons is more efficient than the nonlinear wave-wave
interaction in the short-wavelength KAW range.

In the present paper we study the influence of the observed kinetic-scale
turbulence on the evolution of the proton VDF. First we derive the
Fokker-Planck diffusion terms for protons under the action of Alfv\'{e}nic
turbulence. We are interested here in the influence of the KAW component,
which is dominant in the kinetic-scale SW turbulence, on the evolution of
the proton VDF. The Fokker-Planck diffusion coefficients are presented in
terms of quantities measured in-situ in the solar wind. In particular we
refer to magnetic field measurements that provided spectral indexes of the
SW turbulence at proton kinetic scales \cite{Alexandrova2009,Sahraoui2010},
MHD/kinetic spectral break wavenumber, and the turbulence amplitude at the
break wavenumber \cite{Smith2006, Markovskii2008}. Analytical and numerical
estimations are given for the evolution of nonthermal tails in the proton
VDFs.

\section{Wave Model}

The classic KAW dispersion can be presented as
\begin{equation}
\omega _{k}=k_{\parallel}V_{\rm A}K,  \label{disp}
\end{equation}%
where $\mathbf{k}$ is the wave vector, $\parallel (\perp)$) mean components
parallel(perpendicular)to $\mathbf{B}_{0}$,
$K=K\left( \mu \right) $ is the KAW dispersion function, and $\mu =\rho
_{\rm p}k_{\perp }$ is the dimensionless perpendicular wavenumber.

The kinetic dispersion function for KAWs was derived by \inlinecite{Hasegawa1976}:
\begin{equation}
K=\mu \sqrt{\frac{1}{1-\Lambda _{0}}+\frac{T_{{\rm e}\parallel}}{T_{{\rm p}\perp }}},  \label{K}
\end{equation}%
where $\Lambda _{0}=\Lambda _{0}\left( \mu \right) =I_{0}\left( \mu
^{2}\right) \exp \left( -\mu ^{2}\right) $, $I_{0}\left( \mu ^{2}\right) $
is the zero-order modified Bessel function, and $T_{\rm e(p)}$ is the electron(proton) temperature.
This expression was obtained in
the following wave and plasma parameter ranges: $k_{\parallel}^{2}/k_{\perp }^{2}\ll
1$, $\omega _{k}^{2}/\Omega _{\rm p}^{2}\ll 1$, $V_{{\rm Tp}}^{2}\ll V_{\rm A}^{2}\lesssim
\left( \omega _{k}/k_{\parallel}\right) ^{2}\ll V_{\rm Te}^{2}$.

The KAW dispersion function gives the KAW phase velocity, $V_{k}=\omega
_{k}/k_{\parallel}$, in units of Alfv\'{e}n velocity: $V_{k}/V_{\rm A}=K\left( \mu
\right) $. In the MHD limit $\mu \rightarrow 0$ KAWs become dispersiveless
Alfv\'{e}n waves, $V_{k}\rightarrow V_{\rm A}$, but with growing $\mu $ the KAW
phase velocity deviates from the Alfv\'{e}n velocity significantly. The
inverse function $K^{-1}\left( V_{k}/V_{\rm A}\right) $, which we will need
later in the diffusion coefficient, is impossible to find from Equation (\ref{K})
analytically in general case. Explicit expressions for $K^{-1}\left(
V_{k}/V_{\rm A}\right) $ can only be found in two asymptotic regimes of weak ($%
\mu ^{2}\ll 1$) and strong ($\mu ^{2}\gg 1$) wave dispersion.

To simplify the problem, one can use a Pad\'{e} approximation for $\Lambda
_{0}\left( \mu \right) $,
\begin{equation}
\Lambda _{0}\left( \mu \right) \simeq \frac{1}{1+\mu ^{2}},  \label{Pade}
\end{equation}%
resulting in the following KAW dispersion function
\begin{equation}
K\simeq K_{\mathrm{P}}=\sqrt{1+\left( 1+\frac{T_{{\rm e}\parallel}}{T_{{\rm p}\perp }}\right)
\mu ^{2}}  \label{KP}
\end{equation}%
and its derivative
\begin{equation}
\partial K_{\mathrm{P}}/\partial \mu =\left( 1+\frac{T_{{\rm e}\parallel}}{T_{{\rm p}\perp }}%
\right) \frac{\mu }{K_{\mathrm{P}}}.  \label{KP1}
\end{equation}

These expressions can also be found in the framework of two-fluid MHD plasma
model and provide a good approximation for KAWs in the whole range of $\mu $
if the plasma beta $\beta =\left( 1+T_{{\rm e}\parallel}/T_{{\rm p}\perp }\right)
V_{{\rm Tp}}^{2}/V_{\rm A}^{2}\ll 1$. As is seen from Figure 1, the difference between
the kinetic KAW dispersion $K\left( \mu \right) $ (solid line) and its' Pad%
\'{e} approximation $K_{\mathrm{P}}\left( \mu \right) $ (dashed line) is
insignificant for $T_{{\rm e}\parallel}/T_{{\rm p}\perp }\approx 1$.

\begin{figure}[tbp]
\caption{Comparison of three models for the KAW dispersion function $K\left(
\protect\mu \right) $: (1) kinetic dispersion by Hasegawa and Chen (1976)
for $\protect\beta \ll 1$ (solid line); (2) Pad\'{e} approximation, that
corresponds to the two-fluid MHD dispersion for $\protect\beta \ll 1$ (dash
line); (3) two-fluid dispersion with finite-$\protect\beta $ effects for $%
\protect\beta _{\parallel }=\protect\beta _{\perp }=0.5$ (dot line) and its
asymptote (bottom solid line). In all cases $T_{{\rm e}\parallel}/T_{{\rm p}\perp }=1$.}
\begin{center}
\includegraphics[width=11.5cm]{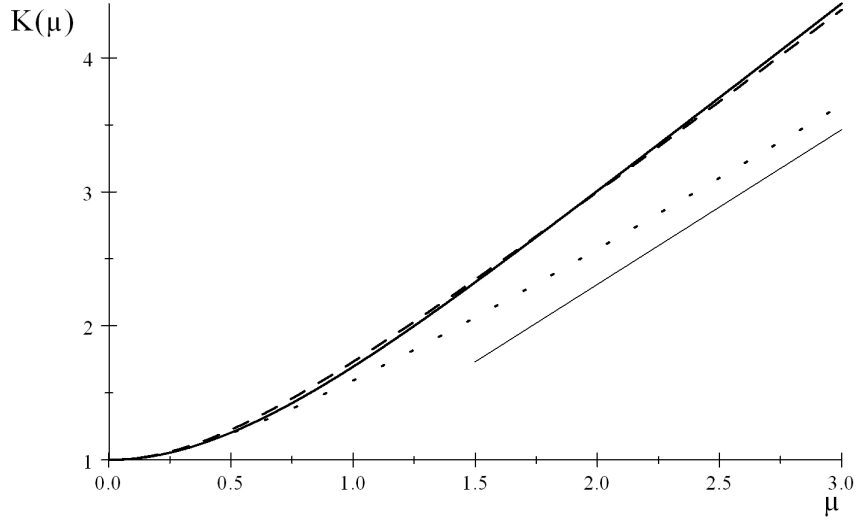}
\end{center}
\label{f1}
\end{figure}

Finite-$\beta $ effects come into play in the solar wind at 1 AU, where the
typical value $\beta \approx 1$ (this corresponds to the plasma/magnetic
pressure ratio $\approx $1). The KAW dispersion function $K_{\beta }$,
accounting for the finite-$\beta $ effects of magnetic and plasma
compressibility \cite{Voitenko2002}, can be modified for the case
of anisotropic temperatures:
\begin{eqnarray}
K_{\beta }^{2} &=&\frac{1+\frac{\beta _{\parallel }}{1+\beta _{\perp }}+%
\frac{1+T_{e\parallel }/T_{p\perp }}{1+\beta _{\perp }}\mu ^{2}}{2}+
\nonumber \\
&&+\sqrt{\left( \frac{1+\frac{\beta _{\parallel }}{1+\beta _{\perp }}+\frac{%
1+T_{e\parallel }/T_{p\perp }}{1+\beta _{\perp }}\mu ^{2}}{2}\right) ^{2}-%
\frac{\beta _{\parallel }}{1+\beta _{\perp }}},  \label{disbeta}
\end{eqnarray}%
where $\beta _{\parallel ,\perp }=\left( 1+T_{{\rm e}\parallel ,\perp
}/T_{{\rm p}\parallel ,\perp }\right) V_{{\rm Tp}\parallel ,\perp }^{2}/V_{\rm A}^{2}$. This
dispersion is also shown in Figure 1 (dot line). The corresponding KAW
wavenumber $\mu$ can be expressed via $V_{k}$ explicitly,
\begin{equation}
\mu =\sqrt{\frac{1+\beta _{\perp }}{1+T_{{\rm e}\parallel}/T_{{\rm p}\perp }}\frac{%
\left( \frac{V_{k}}{V_{\rm A}}\right) ^{4}-\left( 1+\frac{\beta _{\parallel }}{%
1+\beta _{\perp }}\right) \left( \frac{V_{k}}{V_{\rm A}}\right) ^{2}+\frac{\beta
_{\parallel }}{1+\beta _{\perp }}}{\left( \frac{V_{k}}{V_{\rm A}}\right) ^{2}}}.
\label{muR1}
\end{equation}%
At large $\mu $, the KAW dispersion (phase velocity) can be decreased
significantly by the finite-$\beta $ effects. On the contrary, in the weakly
dispersive range $\mu <1$ such moderate values as $\beta =0.5$ do not change
the KAW dispersion significantly and can be approximated there as
\begin{equation}
\mu =K_{\mathrm{P}}^{-1}\left( \frac{V_{k}}{V_{\rm A}}\right) =%
\sqrt{\frac{\left( \frac{V_{k}}{V_{\rm A}}\right) ^{2}-1}{1+\frac{T_{{\rm e}\parallel}}{%
T_{{\rm p}\perp }}}}.  \label{muR}
\end{equation}

\section{Proton Diffusion in the KAW Turbulence}

The parallel component of the KAW electric field $E_{z\mathbf{k}}$ ($%
\mathbf{k}=(k_{x},k_{y},k_{z})$ is the wave vector, $\mathbf{z}\parallel \mathbf{B}_{0}$),
makes KAWs efficient in
Cherenkov interaction with plasma particles. Following \cite{Voitenko1998,Voitenko2003},
we account for the proton VDFs modifications
under the action of KAWs in the framework of quasi-linear theory. We
consider an axially symmetric (with respect to the background magnetic field
$\mathbf{B}_{0}$) problem, where both the KAW spectrum and the particle velocity
distributions are independent of respective polar angles in the cross-field
plane.

In general, the proton distribution function $F=F(\mathbf{r},\mathbf{V},t)$ in the solar wind
obeys the Vlasov collisional equation
\begin{equation}
\left( \frac{\partial }{\partial t}+\mathbf{V\cdot }\frac{\partial }{%
\partial \mathbf{r}}+\mathbf{a}\cdot \frac{\partial }{\partial \mathbf{V}}%
\right) F=\left( \frac{dF}{dt}\right) _{\rm C}+\left( \frac{dF}{dt}\right) _{\rm A},
\label{Vlasov}
\end{equation}%
where $\mathbf{r}$ and $\mathbf{V}$ are respectively the position and velocity vectors of the
protons, $t$ is the time, and $\mathbf{a}$ is the proton acceleration under the
action of the external forces: the macroscopic electric force $Ze\mathbf{E}_{0}$, the
gravitational force $m\mathbf{g}$ and the Lorentz force $\propto \mathbf{v}\times \mathbf{B}_{0}$. The
right-hand side represents the velocity-space proton diffusion due to
Coulomb collisions, $\left( dF/dt\right) _{\rm C}$, and due to wave-particle
collisions,
\[
\left( \frac{dF}{dt}\right) _{\rm A}=\frac{\partial }{\partial V_{z}}D^{\rm A}\frac{%
\partial F}{\partial V_{z}}.
\]

In the diffusion coefficient $D^{\rm A}$ we integrate over the polar angles in
the cross-$\mathbf{B}_{0}$ plain, implying that $\left\vert E_{z\mathbf{k}%
}\right\vert ^{2}$ is axially symmetric, but we keep the dependence on the
perpendicular component of the particle velocity:
\begin{equation}
D^{\rm A}=\frac{\pi q_{\rm p}^{2}}{2m_{\rm p}^{2}}\sum_{\mathbf{k}}\delta \left( \omega
_{k}-k_{z}V_{z}\right) J_{0}^{2}\left\vert E_{z\mathbf{k}}\right\vert ^{2}
\label{D}
\end{equation}%
where the zero-order Bessel function $J_{0}=J_{0}\left( k_{\perp }\rho
_{\rm p}V_{\perp }/V_{{\rm Tp}}\right) $ reflects the fact that the actual electric
field of KAWs acting on the protons is reduced because of the averaging over
their cyclotron orbits with the velocity-dependent gyroradius $V_{\perp
}/\Omega _{\rm p}=\rho _{\rm p}V_{\perp }/V_{{\rm Tp}}$. Other notations are as follows: $%
k_{z}$ and $k_{\perp }$ are the wave vector components parallel and
perpendicular to $B_{0}$, $k_{\perp }^{2}=k_{x}^{2}+k_{y}^{2}$, $\rho
_{\rm p}=V_{{\rm Tp}}/\Omega _{\rm p}$ is the proton gyroradius, $V_{{\rm Tp}}=\sqrt{T_{\rm p}/m_{\rm p}}
$ thermal velocity, $T_{\rm p}$ temperature, $m_{\rm p}$ proton mass, $q_{\rm p}$ proton
charge, and $\sum_{\mathbf{k}}=2\pi \int dk_{z}\int dk_{\perp }k_{\perp }$.
Dirac's delta-function $\delta \left( \omega _{k}-k_{z}V_{z}\right) $
follows from the resonant character of the Cherenkov wave-particle
interaction.

The simplest solar wind models consider only the effects of the external
forces. In such models called exospheric, the right-hand side interaction
terms in Equation (\ref{Vlasov}) are neglected so that an analytic solution of the
equation can be obtained (\inlinecite{Lamy2003} and reference therein). Coulomb
collisions make equations too complex and require numerical simulations.
Expressions for the Coulomb diffusion term in the Vlasov equation for
protons and results of numerical simulations accounting for Coulomb
collisions are presented in \cite{PiVo2012}.

Starting from Equation (\ref{D}), the wave-particle diffusion term accounting for
turbulence properties is derived in next sections. We especially pay
attention to a non-monotonous dependence of the diffusion coefficient on the
parallel velocity, which follows from the non-monotonous dependence of the
spectra of parallel electric fields on the KAW perpendicular wavenumber.

\subsection{Fokker-Planck Diffusion Coefficient in Terms of KAWs' Magnetic
Fields}

Since the parallel electric fields $E_{z\mathbf{k}}$ in KAWs are relatively
weak and difficult to measure, it is instructive to express $D^{\rm A}$ in terms
of KAWs' magnetic fields. To this end one can use the KAW polarization
relation
\begin{equation}
\frac{E_{z}}{B_{\perp }}=-\frac{T_{{\rm e}z}}{T_{{\rm p}\perp }}\frac{V_{\rm A}}{c}\frac{%
k_{z}}{k_{\perp }}\frac{\mu ^{2}}{K}.  \label{Pol}
\end{equation}%
Then the 2D velocity-space diffusion coefficient for the protons can be
reduced to the following integral in the normalized wavenumber space:
\begin{eqnarray}
D^{\rm A} &=&\pi ^{2}\frac{T_{{\rm e}z}}{T_{{\rm p}\perp }}\Omega _{\rm p}V_{\rm S}^{2}\int d\mu
\delta \left( \mu -K^{-1}\right) \frac{\mu ^{3}J_{0}^{2}}{K^{2}\partial
K/\partial \mu }\frac{\int d\nu \nu \left\vert B_{\mu \nu }\right\vert ^{2}}{%
B_{0}^{2}}  \nonumber \\
&=&\pi ^{2}\frac{T_{{\rm e}z}}{T_{{\rm p}\perp }}\Omega _{\rm p}V_{\rm S}^{2}\left[ \frac{\mu
^{3}J_{0}^{2}}{K^{2}\partial K/\partial \mu }\frac{\int d\nu \nu \left\vert
B_{\mu \nu }\right\vert ^{2}}{B_{0}^{2}}\right] _{\mu =K^{-1}}.  \label{D22}
\end{eqnarray}%
The dimensionless parallel wavenumber $\nu =\delta _{\rm p}k_{z}$ ($\delta _{\rm p}$
is the ion inertial length) is introduced in Equation (\ref{D22}), and the following
relation for the Dirac $\delta $-function is used:
\[
\delta \left( K-V_{z}/V_{\rm A}\right) =\frac{\delta \left( \mu -K^{-1}\right) }{%
\partial K/\partial \mu },
\]%
where $K^{-1}=K^{-1}\left( V_{z}/V_{\rm A}\right) $ is the inverse $K$-function
of $V_{z}/V_{\rm A}$ and $V_{\rm S}=\sqrt{T_{{\rm e}z}/m_{\rm p}}$ is the ion-sound speed. The
normalized spectral density, $\left\vert B_{\mu \nu }\right\vert ^{2}=\delta
_{\rm p}^{-1}\rho _{\rm p}^{-2}\left\vert B_{\perp \mathbf{k}}\right\vert ^{2}$, is
defined such that $\int dk_{z}\int dk_{\perp }k_{\perp }\left\vert B_{\perp
\mathbf{k}}\right\vert ^{2}=\int d\nu \int d\mu \mu \left\vert B_{\mu \nu
}\right\vert ^{2}$. Note that $\left\vert B_{\mu \nu }\right\vert ^{2}$ has
the same dimension as $B_{0}^{2}$.

Now we have to express the integral $\int_{0}^{\infty }d\nu \nu \left\vert
B_{\mu \nu }\right\vert ^{2}$ in terms of the integral power at $\mu $, $%
\left\vert B_{\mu }\right\vert ^{2}=\int d\nu \left\vert B_{\mu \nu
}\right\vert ^{2}$. This last quantity is related to the unidirectional
spectral density $W_{\mu }$ measured by spacecraft: $\mu \left\vert B_{\mu
}\right\vert ^{2}=W_{\mu }$. The perpendicular wavenumber $\mu $ is related
to the measured spacecraft-frame frequency $f$ through $\mu \sin \theta
_{VB}\simeq $ $\left( V_{{\rm Tp}}/V_{\rm SW}\right) \left( f/f_{\rm p}\right) $, $\theta
_{VB}$ is the angle between the solar wind velocity $V_{\rm SW}$ and $B_{0}$.

In principle, an unknown spectrum of parallel wavenumbers $\nu $ contributes
to the turbulence level at every particular $\mu $. To simplify the problem,
we take into account the following two facts. First, theory and observations
suggest that the turbulence cascade proceeds along a path in wavenumber
space which is defined by the "critical balance" between the linear and
nonlinear time scales \cite{Goldreich1995}. In accordance to the
critical balance condition, every particular $\mu $ has its own $\nu _{\mu }$
where most of the spectral density is concentrated: $\left\vert B_{\mathbf{%
\mu }\nu }\right\vert ^{2}=\left\vert B_{\mu }\right\vert ^{2}\delta \left(
\nu -\nu _{\mu }\right) $, where $\left\vert B_{\mu }\right\vert ^{2}$ is
the turbulence level at the perpendicular wavenumber $\mu $. Second, the
turbulence fluctuations are anisotropic with $k_{\perp }\gg k_{z}$. Denoting
the anisotropy factor $k_{z}/k_{\perp }\equiv \alpha \left( k_{\perp
}\right) $, the critical balance reads $\nu _{\mu }=\frac{V_{\rm A}}{V_{{\rm Tp}}}%
\alpha \left( \mu \right) \mu $. Hence we estimate $\int_{0}^{\infty }d\nu
\nu \left\vert B_{\mu \nu }\right\vert ^{2}\simeq \frac{V_{\rm A}}{V_{{\rm Tp}}}\alpha
\left( \mu \right) \mu \left\vert B_{\mu }\right\vert ^{2}=$ $\frac{V_{\rm A}}{%
V_{{\rm Tp}}}\alpha \left( \mu \right) W_{\mu }$, and%
\begin{equation}
D^{\rm A}=\pi ^{2}\left( \frac{T_{{\rm e}z}}{T_{{\rm p}\perp }}\right) ^{3/2}\Omega
_{\rm p}V_{\rm S}V_{\rm A}\left[ \frac{\alpha \left( \mu \right) \mu ^{3}J_{0}^{2}}{%
K^{2}\partial K/\partial \mu }\frac{W_{\mu }}{B_{0}^{2}}\right] _{\mu =\mu _{%
\mathrm{v}}},  \label{DD}
\end{equation}%
where $\mu _{\mathrm{v}}=K^{-1}\left( V_{z}/V_{\rm A}\right) $.

\subsection{Diffusion Coefficient for Double-kink Power-law Turbulence
Spectra}

Further progress is possible if we know a particular shape of the turbulent
spectrum$\ $of the Alfv\'{e}nic turbulence. Numerous in-situ observations
indicate that the power law spectra $W_{\mu }\propto \mu ^{-p}$ are typical
in the solar wind, but the spectral index $p$ is different in different
ranges of $\mu $. In the MHD turbulence range $\mu \leq \mu _{1}$ the power
law index $p=p_{1}\simeq 1.7$, close to the Kolmogorov value. In the
intermediate weakly/mildy dispersive KAW range $\mu _{1}\leq \mu \leq 1$ the
spectra are much steeper, with $p=p_{2}$ varying from $p_{2}=2$ to $p_{2}=4$
(Leamon {\it et al.} 1999; Smith {\it et al.} 2011; Sahraoui {\it et al.}, 2010). In the
strongly dispersive KAW range, $\mu \gg 1$, the power index approaches the
value $p=p_{3}\simeq 2.8$ \cite{Alexandrova2009,Sahraoui2010}.
All the above double-kink behavior of $p\left( \mu _{\mathrm{v}}\right) $
can be modeled by the following piecewise function:
\begin{equation}
p=p\left( \mu \right) =p_{1}+\left( p_{2}-p_{1}\right) H\left( \mu -\mu
_{1}\right) +\left( p_{3}-p_{2}\right) H\left( \mu -\mu _{2}\right) ,
\label{d}
\end{equation}%
where $H\left( x\right) $ is the Heaviside step function, $H\left( x\right)
=0$ for $x<0$ and $H\left( x\right) =1$ for $x\geq 0$. The spectral break
(more precisely, the first spectral kink) at $\mu _{1}=0.1 - 0.5$
separates MHD and (weakly/ dispersive) kinetic ranges of Alfv\'{e}nic
turbulence. The second spectral kink, that occurs at $\mu _{2}=2 - 3$ (see
spectra measured by \inlinecite{Sahraoui2010}, separates weakly and strongly
dispersive KAW ranges \cite{Voitenko2011}.

With Equation (\ref{d}), the measured spectral density can be written in the
following general form:
\begin{equation}
W_{\mu }=W_{\mu 1}\left( \frac{\mu _{2}}{\mu _{1}}\right) ^{\left(
p_{3}-p_{2}\right) H\left( \mu -\mu _{2}\right) }\left( \frac{\mu }{\mu _{1}}%
\right) ^{-\left( p_{1}+\left( p_{2}-p_{1}\right) H\left( \mu -\mu
_{1}\right) +\left( p_{3}-p_{2}\right) H\left( \mu -\mu _{2}\right) \right)
},  \label{Wk}
\end{equation}%
where $W_{\mu 1}$ is the turbulence spectral density at $\mu =\mu _{1}$
(note that $W_{\mu 1}$ has dimension of $B_{0}^{2}$). The re-normalization
of the spectrum (Equation (\ref{Wk})) at $\mu \geq \mu _{2}$ ensures it is continuous
through the second kink at $\mu =\mu _{2}$.

With Equation (\ref{Wk}) the velocity diffusion coefficient (Equation (\ref{DD})) attains the
following general form:
\begin{eqnarray}
D^{\mathrm{A}} &=&\pi ^{2}\Omega _{\mathrm{p}}V_{\mathrm{S}}V_{\mathrm{A}}%
\frac{\left( \frac{T_{\mathrm{e}z}}{T_{\mathrm{p}\perp }}\right) ^{3/2}}{%
\left( 1+\frac{T_{\mathrm{e}z}}{T_{\mathrm{p}\perp }}\right) }\times
\nonumber \\
&&\times \frac{\mu _{1}^{2}\alpha \left( \mu _{\mathrm{v}}\right) \left(
\frac{\mu _{2}}{\mu _{1}}\right) ^{\left( p_{3}-p_{2}\right) H\left( \mu _{%
\mathrm{v}}-\mu _{2}\right) }\left( \frac{\mu _{\mathrm{v}}}{\mu _{1}}%
\right) ^{2-p\left( \mu _{\mathrm{v}}\right) }J_{0}^{2}\left( \mu _{\mathrm{v%
}}\frac{V_{\perp }}{V_{\mathrm{Tp}}}\right) }{\frac{V_{z}}{V_{\mathrm{A}}}}%
\frac{W_{\mu 1}}{B_{0}^{2}},  \label{D2}
\end{eqnarray}%
where $\mu _{\mathrm{v}}$= $\mu (V_{k}=V_{\parallel})$ is the resonant KAW wavenumber
defined by Equation (\ref{muR1}) and $p\left( \mu _{\mathrm{v}}\right) $ is given by
Equation (\ref{d}). In what follows we will put $\alpha
\left( \mu \right) \simeq 0.1$ at ion kinetic scales $\mu \approx 1$
\cite{Sahraoui2010}, and neglect its relatively slow variation
in the resonant wavenumber range.

The diffusion due to KAWs is strongest for the protons with velocities
around $V_{z}\simeq V_{1}\gtrsim V_{\rm A}$, which are Cherenkov-resonant with
the spectral peak of the KAW parallel electric fields. In the plane $\perp
B_{0}$, the diffusion is maximized for the core protons at $V_{\perp
}\lesssim V_{{\rm Tp}}$, where the reduction due to proton-cyclotron gyration is
minimized. In the $B_{0}$-parallel direction, the diffusion coefficient
decreases very fast as $V_{z}\rightarrow V_{\rm A}$, and less fast with $V_{z}$
growing in a more extended velocity range from $V_{1}$ to several Alfv\'{e}n
velocities. These features are compatible with the physical picture of the
Cherenkov resonant interaction between gyrating particles and waves with
finite cross-$B_{0}$ length scales.

Since there are no positive $\mu =K^{-1}\left( V_{z}/V_{\rm A}\right) $ for $%
V_{z}/V_{\rm A}<1$, $D^{\rm A}=0$ there (this fact reflects the absence of resonant
KAWs for particles moving with sub-Alfv\'{e}nic velocities $V_{z}/V_{\rm A}<1$).
One should note, however, that even sub-Alfv\'{e}nic protons can be affected
by the turbulence fluctuations via two effects: (i) nonlinear broadening of
the Cherenkov resonances, and (ii) increased Coulomb diffusion of the
protons at the steep VDF slope at $V_{z}\lesssim V_{\rm A}$. The minimum
velocity of the affected protons resulting from the former effect is
analytically estimated below in Equation (\ref{Vmin}).

\begin{figure}[tbp]
\caption{Reduced diffusion coefficient as function of normalized parallel
velocity $V=V_{z}/V_{\rm A}$
for $p_{2}=2$ (dot line), $p_{2}=3$
(dash line), and $p_{2}=4$ (solid line). Turbulence level at the first
spectral kink is adjusted for each $p_{2}$ using the scaling $W_{\protect\mu %
1}\propto p^{7.4}$. }
\begin{center}
\includegraphics[width=11.5cm]{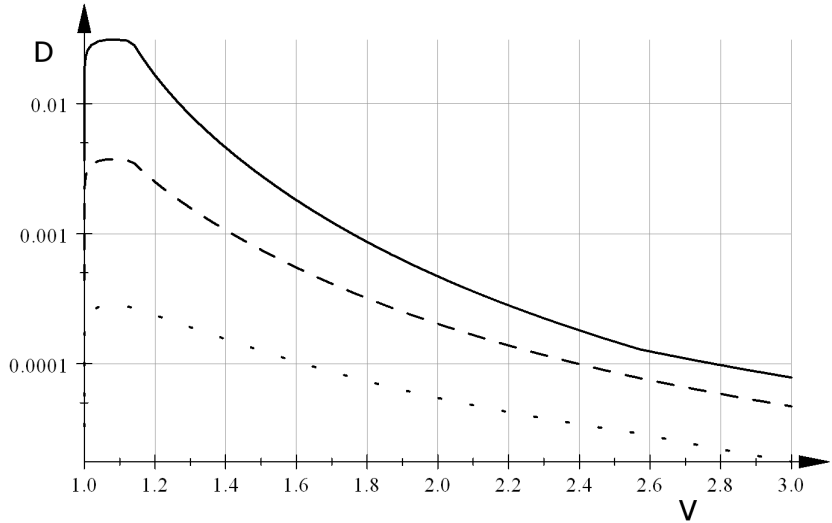}
\end{center}
\label{fig2}
\end{figure}

It is interesting to note that the spectral slope $p\left( \mu _{\mathrm{v}%
}\right) <2$ in (\ref{d}) can make $D^{\rm A}$ a still increasing function of $%
\mu _{\mathrm{v}}$ in the dissipation range between $\mu _{1}$ and $\mu _{2}$%
. This follows from the increase of the parallel electric fields in KAWs
that is faster than the power-law decrease of the turbulence amplitudes.
Therefore, depending on $p\left( \mu _{\mathrm{v}}\right) \lessgtr 2$, the
diffusion coefficient attains a maximum value at $\mu _{\mathrm{v}}=\mu _{1}$
(with $p\left( \mu _{\mathrm{v}}\right) >2$) or at $\mu _{\mathrm{v}%
}\lesssim \mu _{2}$ (with $p\left( \mu _{\mathrm{v}}\right) <2$). Such
behavior of the diffusion coefficient could have an important consequence
that the less steep dissipation-range spectra result in longer nonthermal
tails in the proton VDFs. However this could only be true for the cases
where the turbulence amplitude is the same at $\mu =\mu _{1}$.

In fact, as was found by \inlinecite{Smith2006} in the inertial range, the
larger spectral fluxes $\epsilon $ are followed by the steeper
dissipation-range spectra $p_{2}$, such that $p_{2}=1.05\epsilon ^{0.09}$.
Applying this relation at the end of inertial range, we find the scaling
\begin{equation}
W_{\mu 1}\propto p_{2}^{7.4},  \label{scaling}
\end{equation}%
relating the spectral power $W_{\mu 1}$ at the break $\mu =\mu _{1}$ and the
spectral index $p_{2}$ above it. If the spectral powers in two cases $\left(
1\right) $ and $\left( 2\right) $ are $W_{\mu 1}^{\left( 1\right) }$ and $%
W_{\mu 1}^{\left( 2\right) }$, then the ratio $W_{\mu 1}^{\left( 1\right)
}/W_{\mu 1}^{\left( 2\right) }=$ $\left( p_{2}^{\left( 1\right)
}/p_{2}^{\left( 2\right) }\right) ^{7.4}$. Elaborating this scaling
relation, a reduced 1D diffusion coefficient is shown in Figure 2 for three
spectral indices $p_{2}=4,$ $3$, and $2$ from top to bottom. The reduction
was performed by the averaging of $J_{0}^{2}$ in (\ref{D2}) in the
cross-field velocity plane. With the correlation found by Smith {\it et al.}
(2006), the diffusion coefficient, in average, is larger for steeper
dissipation-range spectra, even if they decrease with $V_{z}$ faster.

The described above properties of the diffusion coefficient indicate that
kinetic Alfv\'{e}n turbulence can be efficient in diffusing the protons in
the velocity range covered by nonthermal tails of the proton VDFs. This is
supported by the analytical and numerical estimations presented below.

\section{Evolution of Nonthermal Tails in the Proton VDFs}

We investigate here what nonthermal proton tails can be formed by their
quasilinear diffusion along the kinetic turbulent spectrum described above.
Let us consider a local evolution problem in the solar wind neglecting external forces,
spatial variations, and Coulomb collisions in Equation (\ref{Vlasov}). To estimate
the formation time for a particular tail, we have to know the tail extension
in the field-aligned direction $V_{\max }$.
Observations show that the tail extension $V_{\max }$ rarely exceeds $3V_{\rm A}$,
which allows us to simplify the problem accounting only for the
intermediate KAW spectrum in the diffusion coefficient (Equation (\ref{D2})). The
cross-field velocity spread in the tails is approximately the same as in the
proton core, $\approx V_{{\rm Tp}\perp }$ (within the multiplier of the order 1).

Accounting for above, we reduce Equation (\ref{Vlasov}) to the 1D diffusion equation
\begin{equation}
\frac{\partial f}{\partial t}=\frac{\partial }{\partial V_{z}}D\frac{%
\partial f}{\partial V_{z}}  \label{Deq}
\end{equation}%
for the reduced 1D distribution function $f=f\left( V_{z}\right) =$ $2\pi
\int dV_{\perp }V_{\perp }F\left( V_{z},V_{\perp }\right) $. The 1D
diffusion coefficient in the range $V_{\rm A}<V_{z}<3V_{\rm A}$,
\begin{equation}
D\simeq 0.1\pi ^{2}\Omega _{\rm p}V_{\rm S}V_{\rm A}\frac{\left( T_{{\rm e}z}/T_{{\rm p}\perp
}\right) ^{3/2}}{\left( 1+T_{{\rm e}z}/T_{{\rm p}\perp }\right) }\frac{\mu
_{1}^{p_{2}}\Lambda _{0}\left( \mu _{\mathrm{v}}^{2}\right) }{\mu _{\mathrm{v%
}}^{p_{2}-2}}\frac{V_{\rm A}}{V_{z}}\frac{W_{\mu 1}}{B_{0}^{2}},  \label{D3}
\end{equation}%
follows from Equation (\ref{D2}) after averaging over Maxwellian distribution in the
cross-field plane: $\left\langle J_{0}^{2}\left( \mu _{\mathrm{v}}\frac{%
V_{\perp }}{V_{{\rm Tp}}}\right) \right\rangle =\Lambda _{0}\left( \mu _{\mathrm{v}%
}^{2}\right) $.

For small $\mu _{1}$ the corresponding resonant velocity $V_{1}=V_{\rm A}K\left(
\mu _{1}\right) $ is close to $V_{\rm A}$ and the remaining velocity interval $%
V_{\rm A}<V_{z}<V_{1}$ is narrow. The detailed behavior of the diffusion
coefficient $D$ is unimportant there because $D$ decreases fast to $0$ with $%
V_{z}\rightarrow V_{\rm A}$, and the nonlinear resonance broadening of KAWs with
$\mu \simeq \mu _{1}$ is more important.

\subsection{Quasilinear Plateau Model}

In this section we study the quasilinear plateau evolution under the
influence of the KAW turbulence. To this end we start from the initially
isotropic Maxwellian VDF for the protons
\[
f_{t=0}=f_{\rm M}=\frac{1}{\sqrt{2\pi }V_{{\rm Tp}\parallel }}\exp \left( -\frac{%
V_{z}^{2}}{2V_{{\rm Tp}\parallel }^{2}}\right) .
\]%
The diffusion coefficient (Equation (\ref{D3})) and hence the velocity-space diffusion
are very non-uniform, as is seen from Figure 2.
The diffusion is fastest in the vicinity of $V_{\rm A}$, where the
diffusion coefficient is maximal, but decreases about two orders at $3V_{\rm A}$.
The quasilinear saturation in this situation occurs first around this maximum,
where an initial local plateau is formed as is shown in Figure 3 (dashed line).
The resulting 1D proton VDF can be modeled
by the piece-wise function with the local plateau
\begin{equation}
f=f_{\rm pl}=\frac{\bar{n}_{\rm pl}}{V_{\max }-V_{\min }}  \label{fpl}
\end{equation}
in the velocity range $V_{\min }<V_{z}<V_{\max }$, and $f$ remaining
Maxwellian,
$f= f_{\rm M}$, outside the plateau.

\begin{figure}[tbp]
\caption{Model proton VDF with a step-like quasilinear plateau extending to $%
V_{\max }=$ $2.5V_{\rm A}$ at the late evolutional stage (an early evolutional
stage is shown by dash line). Numerical solution of the diffusion equation
(Equation (\ref{Deq})) for the same time is shown by stars. A good
correspondence is observed between the model and numerically calculated
VDFs. Details on the plasma and turbulence parameters used are given in the
text.}
\begin{center}
\includegraphics[width=11.5cm]{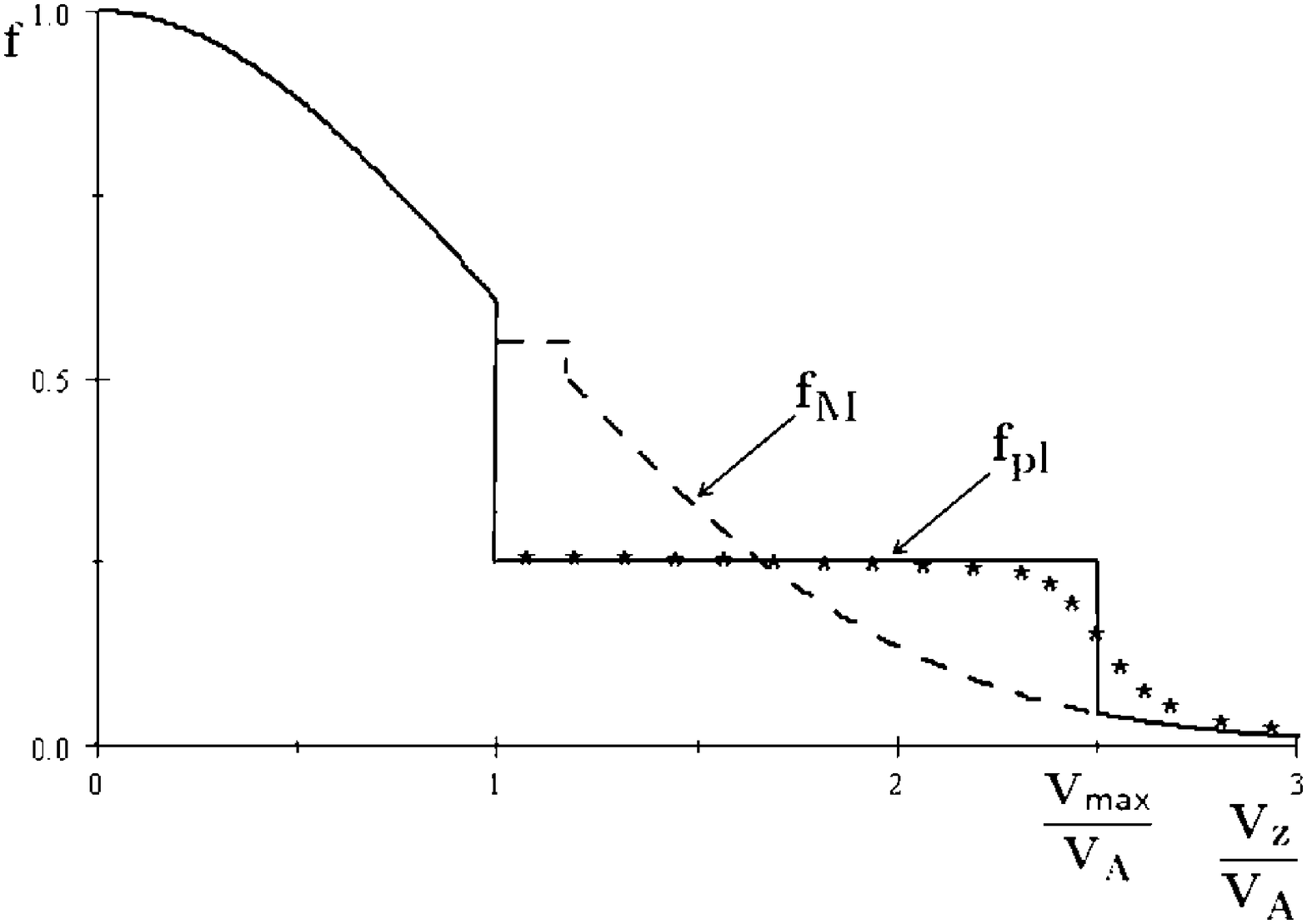}
\end{center}
\label{fig3}
\end{figure}

The quasilinear saturation spreads in time to higher velocities,
shifting the plateau front $V_{\max }=V_{\max }\left( t\right) $ forward to higher
velocities as is also shown in Figure 3 (solid line).
The normalized proton number density in the plateau can be expressed in terms of
error function Erf:
\begin{equation}
\bar{n}_{\rm pl}=\frac{n_{\rm pl}}{n_{0}}=\int_{V_{\min }}^{V_{\max }}dV_{z}f_{\rm M}=%
\frac{\mathrm{Erf}\left( \frac{V_{\max }}{\sqrt{2}V_{{\rm Tp}}}\right) -\mathrm{Erf}%
\left( \frac{V_{\min }}{\sqrt{2}V_{{\rm Tp}}}\right) }{2}.  \label{npl}
\end{equation}%
It increases with time, but the plateau height decreases.

It is important to note that the minimum velocity of the protons involved in
the plateau formation, $V_{\min }$, can be less than the minimum phase
velocity of resonant KAWs, $V_{\rm A}$. As follows from the nonlinear broadening
of the Cherenkov resonance for a KAW with wavenumber $k_{\perp }$ and
amplitude $B_{k}$, the minimum velocity of the affected protons is
\begin{equation}
\frac{V_{\min }}{V_{\rm A}}=\sqrt{1+k_{\perp }^{2}\rho _{{\rm T}}^{2}}-\sqrt{4\frac{%
V_{{\rm p}\parallel }}{V_{\rm A}}\frac{T_{{\rm e}\parallel }/T_{{\rm p}\parallel }}{\sqrt{%
T_{{\rm p}\perp }/T_{{\rm p}\parallel }+T_{{\rm e}\parallel }/T_{{\rm p}\parallel }}}}\sqrt{\frac{%
k_{\perp }\rho _{{\rm T}}}{\sqrt{1+k_{\perp }^{2}\rho _{{\rm T}}^{2}}}\frac{B_{k}}{B_{0}}%
}.  \label{Vmin}
\end{equation}%
This formula follows from the condition that the proton kinetic energy in
the wave frame is equal to the KAW potential barrier. For $T_{{\rm e}\parallel
}/T_{{\rm p}\parallel }=$ $T_{{\rm p}\perp }/T_{{\rm p}\parallel }=1$, and the typical KAW
amplitude $B_{k}/B_{0}=0.05$ in the vicinity of the first spectral break, at
$k_{\perp }\rho _{\rm T}=\sqrt{1+T_{{\rm e}\parallel }/T_{{\rm p}\perp }}k_{\perp }\rho
_{\rm p}\simeq 0.3$, the minimum velocity reduces to $V_{\min }\simeq 0.8V_{\rm A}$.
The modified $V_{\min }<V_{\rm A}$ can significantly increase the proton number
density in the tail, and hence the energy in the parallel degree of freedom
in comparison to the case $V_{\min }=V_{\rm A}$. On the other hand, such
decrease of $V_{\min }$ does not influence much the evolution time scales of
the tails.

\subsection{Evolution Time Scale of the Quasilinear Plateau}

The time evolution of the plateau spreading can be found from the diffusion
equation (Equation (\ref{Deq})) as follows. First, we note that the time derivative of $%
f_{\rm pl}$ can be expressed as
\begin{equation}
\frac{\partial f_{\rm pl}}{\partial t}=-\frac{f_{\rm pl}-f_{\rm M}\left( V_{\max
}\right) }{V_{\max }-V_{\min }}\frac{\partial V_{\max }}{\partial t}.
\label{dfpl}
\end{equation}%
Using expression (\ref{dfpl}) in the lhs of the diffusion equation (Equation (\ref{Deq})) and then
integrating it over the plateau velocity range, from $V_{\min }+0$ to $V_{\max }+0$,
we obtain a
first-order ordinary differential equation for $V_{\max }$ as function of
time $t$:
\[
\left( f_{\rm pl}-f_{\rm M}\left( V_{\max }\right) \right) \frac{dV_{\max }}{dt}%
=D\left( V_{\max }\right) \frac{\partial f_{\rm M}\left( V_{\max }\right) }{%
\partial V_{\max }}.
\]%
This equation can be solved easily by separating variables and integrating
from $t=0$ to $t$:
\begin{equation}
t=V_{{\rm Tp}\parallel }^{2}\int_{V_{\min }}^{V_{\max }}\frac{f_{\rm pl}\left( V_{\max
}^{\prime }\right) /f_{\rm M}\left( V_{\max }^{\prime }\right) -1}{V_{\max
}^{\prime }D\left( V_{\max }^{\prime }\right) }dV_{\max }^{\prime }.
\label{tV}
\end{equation}

Let us estimate the plateau formation time at 1 AU, where $\Omega _{\rm p}\simeq 1
$ s$^{-1}$, $T_{{\rm e}z}/T_{{\rm p}\perp }\simeq $ $T_{{\rm e}z}/T_{{\rm p}z}\simeq 1$ and $%
V_{\rm A}\simeq V_{{\rm Tp}}\simeq 50$ km s$^{-1}$. With the turbulence amplitude $\left\vert
B_{1}\right\vert /B_{0}\simeq 0.05$, spectral break wavenumber $\mu
_{1}\simeq 0.6$, and spectral slope $p_{2}=3$ just above the break, (\ref{tV}%
) gives the time about half an hour for formation of quasilinear plateau with
the length equal to the average tail length $V_{\max }=1.75V_{\rm A}$. Since
this time is much (more than 2 orders) shorter than the solar wind expansion
time $t_{\mathrm{SW}}$ (2-4 days at 1 AU), such tails can be easily generated by
the turbulence locally in the solar wind.

A long tail with $V_{\max }$ reaching $2.5V_{\rm A}$ is developed by the $p_{2}=2$
turbulence at $t=$ 14 h, when the local approximation is still
marginally applicable. The corresponding model solution is shown in Figure 3
by the solid line. Furthermore, to verify the model, we solve numerically
the proton diffusion equation (Equation (\ref{Deq})) with the zero-gradient
Neumann boundary conditions
at the lower ($V_{z}/V_{\rm A}=1$) and upper ($V_{z}/V_{\rm A}=5$) boundaries. The
value for the upper boundary is chosen in such a way that its variations
does not affect solution in the velocity range $V_{z}/V_{\rm A}=1 - 3$. This
solution is also shown in Figure 3 by stars. A good correspondence between
numerical and analytical solutions indicates that the model
(Equations (\ref{fpl})--(\ref{tV})) is sufficient for quantitative estimations
of the quasilinear plateau height, length, and evolution time-scale.

For longest tails with $V_{\max }\simeq 3V_{\rm A}$, the tail formation time
approaches the solar wind expansion time $t_{\mathrm{SW}}=$ $2 - 4$ days.
This means that the local analysis for such tails is inapplicable, but does
not mean that such tails cannot be formed by the turbulence. To study such
tails one needs to solve a nonlocal problem incorporating radial dependencies
of the solar wind plasma and turbulence characteristics (this work is in
progress). Again, even for such long tails as $V_{\max }\simeq 3V_{\rm A}$, the
problem can be made local with slightly increased turbulence amplitudes
and/or spectral break wavenumbers, say to $\left\vert B_{1}\right\vert
/B_{0}\simeq 0.07$ and $\mu _{1}\simeq 0.8$. Such variations are within the
ranges of measured values (see Figure 2 by \inlinecite{Markovskii2008}. In
addition, as we will see below, the decrease of the spectral slope $p_{2}$
(with fixed amplitude $B_{1}$) has the same effect.

Since the plateau formation time given by Equation (\ref{tV}) is quite complex function
of plasma and turbulence parameters, its radial dependence require further
investigations using particular plasma and turbulence models.
For example,
with approximately constant break frequency (constant $k_{\perp 1}$),
the plateau formation time decreases
with decreasing $R$ as, roughly, $\sim V_{\rm A}\sim R$, the same as $t_{\mathrm{SW}}$.
In this case the locality condition can be still satisfied much closer to the Sun,
where $V_{\rm A}$ is larger and hence longer tails can be generated.
Several numerical estimations with reasonable plasma and
turbulence parameters have shown that the tails
$V_{\max } \approx 2V_{\rm A}$ can be easily
generated, within half an hour, at such short heliocentric distances as $R\approx 0.1$ AU, where
$t_{\mathrm{SW}}\approx 5-7$ h.
The conditions are even more favorable with constant  $\mu _{1}$.
These estimations are compatible with Helios observations
of nonthermal proton tails at all distances from 0.3 to 1 AU.

\begin{figure}[tbp]
\caption{Normalized tail length $V_{\max }/V_{\rm A}$ as function of normalized
time $t/t_{\mathrm{SW}}$. Spectral slopes are $p_{2}=$ 2, 2.5, 3, 3.5, and 4
from the top curve to bottom. All other plasma and turbulence parameters are
the same for all curves (see text). }
\begin{center}
\includegraphics[width=11.5cm]{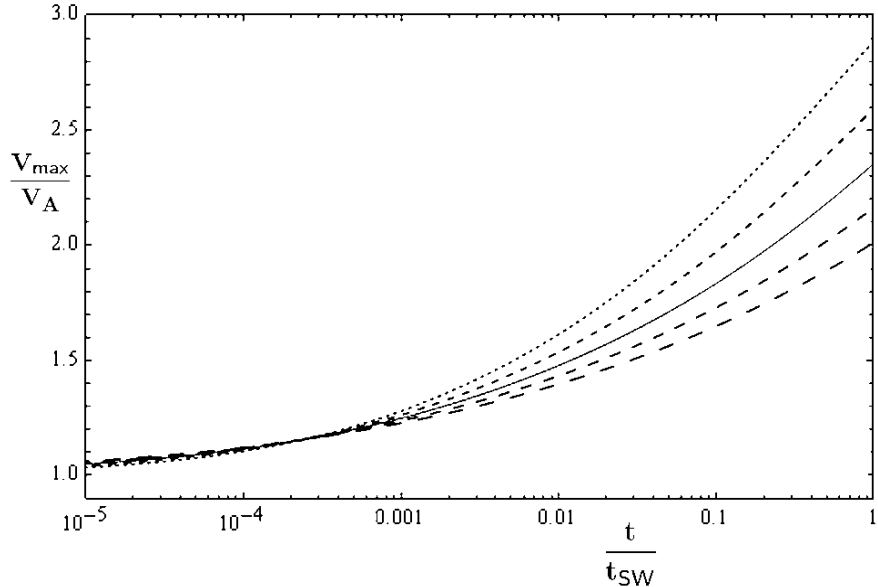}
\end{center}
\label{f4}
\end{figure}

Figure 4 shows the time evolution of the plateau boundary $V_{\max }
$=$V_{\max }\left( t\right) $ in a turbulence with the spectral power $%
W_{f1}\simeq 0.2$ nT$^{2}$ Hz$^{-1}$ at the break frequency $f_{1}\simeq 0.3$ Hz.
These values are well within ranges of measured values and close to the
average ones (see Figures 1 and 2 by \inlinecite{Markovskii2008}. Five curves in
Figure 4 correspond to five different spectral slopes $p_{2}=$2, 2.5, 3, 3.5,
and 4 (from top to bottom). Longer dashes indicate larger $p_{2}$, except
for the solid curve with $p_{2}=$3. It is seen that the formation time $t=t_{%
\mathrm{QL}}$ for the plateau $V_{\max }=$ $\left( 1.5 - 2.5\right) V_{\rm A}$
is considerably shorter than the solar wind expansion time, $t_{\mathrm{QL}%
}/t_{\mathrm{SW}}=$ $0.01 - 0.3$ (we assume an average $t_{\mathrm{SW}%
}=2.7\times 10^{5}$ s). With these values, covering typical proton
nonthermal tails observed in the solar wind, the quasilinear diffusion
process can be considered in the local approximation and our results are
self-consistent. As the tail length approaches $3V_{\rm A}$, the corresponding
evolution time scale approaches $t_{\mathrm{SW}}$ for the shallowest spectra.
This suggests that in average plasma conditions the tail length can hardly
exceed $3V_{\rm A}$, except for the cases of high turbulence levels combined
with shallow spectra.
\begin{figure}[tbp]
\caption{Formation time of the tail with $V_{\max }=$ $1.5V_{\rm A}$ as function
of the proton thermal/Alfv\'{e}n velocity ratio $V_{{\rm Tp}}/V_{\rm A}$ for the
spectral slopes $p_{2}=$ 2.5, 3, 3.5, and 4 (from top to bottom). Other
parameters are as in Figures 3 and 4.}
\begin{center}
\includegraphics[width=11.5cm]{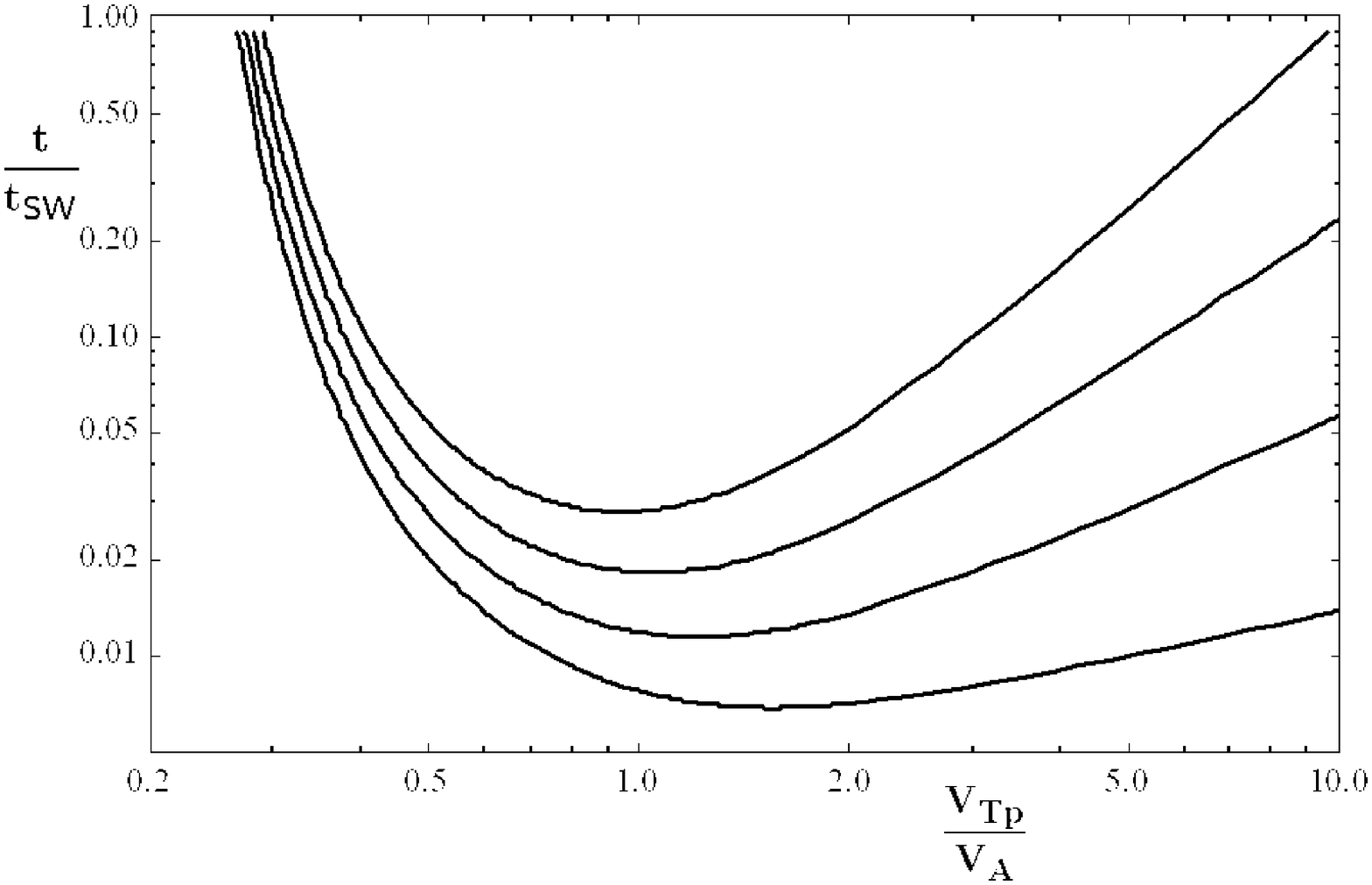}
\end{center}
\label{fig5}
\end{figure}

In Figure 5 we are looking for the favorable conditions facilitating local
production of nonthermal proton tails in the solar wind. To this end we fix
the plateau front at $V_{\max }=1.5$ and plot the corresponding formation
time as function of $V_{{\rm Tp}}/V_{\rm A}$ (which is proportional to the square root
of the proton plasma beta). The spectral indexes are $p_{2}=$2.5, 3, 3.5,
and 4 for the four curves from bottom to top. The most favorable $%
V_{{\rm Tp}}/V_{\rm A}$ for each curve is that one at which $t/t_{\mathrm{SW}}$
attains minimum. The shallowest spectra appeared to be most efficient at $%
V_{{\rm Tp}}/V_{\rm A}\simeq 1.5$, whereas the steepest spectra at $%
V_{{\rm Tp}}/V_{\rm A}\lesssim 1$. These values are not much different for all spectral
slopes of interest, so that we can take the optimal value $V_{{\rm Tp}}/V_{\rm A}\approx 1$,
which is close to the average value in the solar wind at 1 AU. This means
that the turbulence at 1 AU become more efficient in producing
nonthermal tails.

In general, as is seen from Figures 4 and 5, the shallower kinetic spectra are
more efficient in the tail production provided all other turbulence
characteristics are fixed. However, not only the spectral slope, but also
the level of turbulence is highly variable in the solar wind. Moreover, as
was noticed by \inlinecite{Smith2006}, there is a positive correlation between
the spectral flux before the spectral break and the spectral slope after it.
The corresponding scaling in terms of spectral power can be written as $%
W_{\mu 1}\propto p_{2}^{7.4}$. Rescaling the turbulence amplitudes
corresponding to $p_{2}=$2, 2.5, 3.5, and 4 slopes, we plot the resulting
plateau front velocities for each $p_{2}$ in Figure 6. Now we obtain an opposite
ordering of $V_{\max }$ with $p_{2}$, indicating that longer tails are
generated by steeper spectra. Another new feature is less scattered final
values of $V_{\max }$ as compared to the case of uncorrelated
amplitudes and slopes.

\begin{figure}[tbp]
\caption{Normalized tail length $V_{\max }/V_{\rm A}$ as function of normalized
time $t/t_{\mathrm{SW}}$ for the same parameters as in Figure 4 except for the
turbulence amplitudes are now adjusted by the scaling $W_{\protect\mu %
1}\propto p_{2}^{7.4}$. In this case the order of curves is reversed, $p_{2}=
$ 2, 2.5, 3, 3.5, and 4 from bottom to top, which means the longer tails are
generated by steeper spectra. }
\begin{center}
\includegraphics[width=11.5cm]{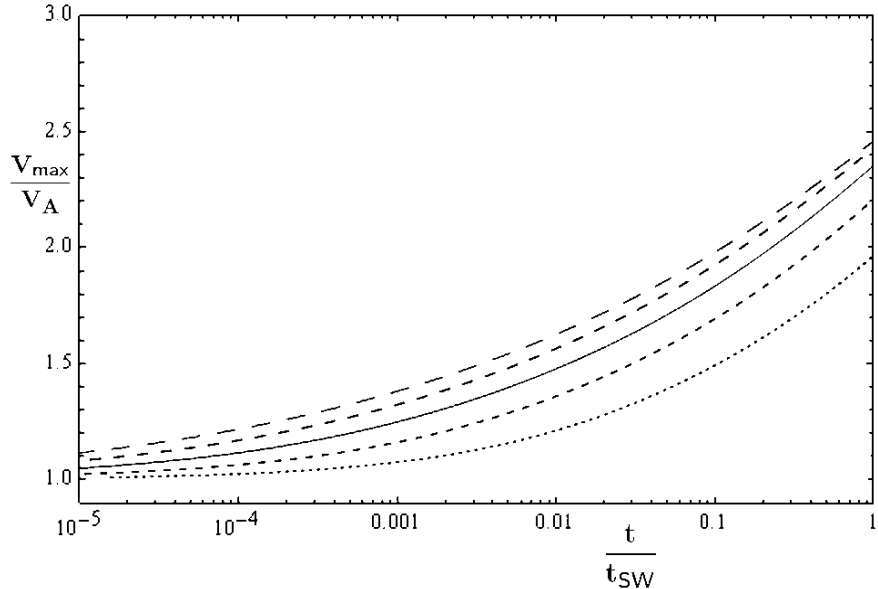}
\end{center}
\label{fig6}
\end{figure}

\subsection{Numerical Example}

At present, there are no data on the proton VDF and kinetic-scale
turbulence measured simultaneously. The problem with the Cluster spacecraft
is that they are capable of measuring the turbulence at sufficient frequency
resolution, but the proton velocity-space resolution is insufficient in
the solar wind conditions. It
is nevertheless interesting to estimate a nonthermal proton tail that
could result from a turbulence measured by Cluster.

In a particular case of Cluster measurements 10 January 2004 from 06:15 to
06:25 UT, reported by Sahraoui {\it et al.} (2010), the background plasma
parameters are: $B_{0}=10.2$ nT, $n_{0}=16$ cm$^{-3}$, $T_{\rm e}=10.4$ eV, $%
T_{\rm p}=31$ eV, $V_{\rm SW}=$548 km s$^{-1}$, angle between the solar wind velocity and
magnetic field $\theta _{BV}=67^{\circ }$. Therefore, the proton thermal
velocity in this case is about the same as the Alfv\'{e}n velocity, $%
V_{{\rm Tp}}\simeq V_{\rm A}\simeq 55$ km s$^{-1}$, proton cyclotron frequency $\Omega
_{\rm p}\simeq 0.94$ s$^{-1}$, proton gyroradius $\rho _{\rm p}\simeq \allowbreak 59$
km, and solar wind expansion time $t_{\mathrm{SW}}=76$ hours. The key
turbulence characteristics are: spectral index at the proton kinetic scales $%
p=p_{2}=4$, spectral break $\mu _{1}=0.4$, and the turbulence spectral
density $W_{\mu 1}\simeq 0.15$ nT$^{2}$ at $\mu =\mu _{1}$.

The dependence $V_{\max }\left( t\right) $ for this case is given by the
bottom line in Figure 4. One can observe that the turbulence should generate a
nonthermal proton tail with $V_{\max }\lesssim 2V_{\rm A}$ at 1 AU. The proton
VDF could not be measured by Cluster, but the average parallel energy of the
protons was found to be larger than the perpendicular one, producing the
effective temperature anisotropy $T_{{\rm p}\parallel }/T_{{\rm p}\perp }\lesssim 1.5$.
This anisotropy cannot be produced by the quasilinear evolution of the
initially isotropic Maxwellian VDF in the velocity range from $V_{\rm A}$ to $2V_{\rm A}$. To
obtain such anisotropy, one needs either more protons in the tail,
or longer tail ($V_{\max }>2V_{\rm A}$), or both.
The additional protons can be delivered to the tail by the nonlinear
broadening of the Cherenkov resonance at $\mu \simeq \mu _{1}$, as is given
by Equation (\ref{Vmin}), and/or by the enhanced collisional diffusion at the
steep velocity-space gradients at $V_{z}\simeq V_{\min }$, where VDF
undergoes a sharp transition from the Maxwellial VDF to the plateau-like. Most
probably, both these processes have non-negligible effects on the tail
density, and act synergetically. Again, we would like to stress that the
tail length is not much affected by these processes. The longer
tail could only be generated at shorter radial distances to the Sun, in more
favorable conditions.

\section{Conclusions and Discussion}

We studied a Fokker-Planck proton diffusion in the solar wind induced by
Alfv\'{e}nic turbulence observed recently at proton kinetic scales
(\opencite{Sahraoui2010}; \opencite{He2011},\citeyear{He2012};
\opencite{Podesta2011}; \opencite{Salem2012}).
Observations indicate that the turbulence is anisotropic and dominated by
large perpendicular wavevectors, which is typical for KAWs possessing
parallel electric fields and hence experiencing Cherenkov resonances. In the
presence of a wide KAW spectrum with overlapping harmonics, the protons
undergo numerous Cherenkov kicks resulting in a Fokker-Planck proton
diffusion in the velocity space along the mean magnetic field.

We derived a quasilinear diffusion equation with the diffusion coefficient
expressed in terms of measured turbulence parameters: spectral break
wavenumber $\mu _{1}$, turbulence amplitude $W_{\mu 1}$ at the spectral
break, and spectral slopes $p_{2}$ below the break. At small and large $\mu $,
the asymptotic values for the spectral index are $p\simeq 1.7$ and $2.7$,
respectively ({\it e.g.} \opencite{Alexandrova2009}). There is however an
intermediate spectral range, $\mu _{1}<\mu <\mu _{2}$, where the spectrum
slope is variable and often steepens to $p\simeq 3 - 4$ or even higher
values \cite{Sahraoui2010}. The turbulence power at these scales is
mostly in form of weakly/mildy dispersive KAWs that are Cherenkov-resonant
with protons in the velocity range covered by observed nonthermal tails,
$V_{z}\simeq \left( 1 - 3\right) V_{\rm A}$. Enhanced nonlinear interaction
among KAWs can explain such steep spectra in this range (Voitenko and De
Keyser, 2011), but resonant generation of nonthermal tails can contribute to
the steepening as well. On the other hand, an additional source for the KAW
replenishing in the dissipation range can be provided by the non-local
nonlinear coupling among KAWs and large-scale MHD AWs \cite{Zhao2011}.

\inlinecite{Rudakov2012} applied an asymptotic
analysis to study the quasilinear proton diffusion in the strongly
dispersive KAW turbulence with $\mu \gg 1$, which implies relatively
high-energy resonant protons with $V_{z}\gg V_{\rm A}$. However, such asymptotical
analysis is inapplicable to typical tails with proton velocities
$V_{\rm A}<V_{z}<3V_{\rm A}$. The reason is that the KAW phase velocity
(Equation (\ref{disbeta}))
cannot be approximated by a simple power-law dependence in the resonant
wavenumber range $0.3<\mu <3$. We focused on this tail-resonant
range, where most of kinetic-scale power is residing and where the full
KAW dispersion (Equatiion (\ref{disbeta})) has to be used. Consequently, the resulting
diffusion equation is more complex and does not have an immediate
analytical solution similar to that found by \inlinecite{Rudakov2012}. To
proceed further analytically, we took into account a fast decrease of the
diffusion coefficient with growing $V_{z}$ (see Figure 2), which allowed us to
formulate a simple yet plausible model for the quasilinear plateau, Equations
(\ref{npl}) and (\ref{npl}), with the time-dependent front velocity $V_{\max }$ and
height $f_{\rm pl}$. By the use of this model in the diffusion equation
(Equations (\ref{Deq}) and (\ref{D3})), we found an analytical solution (Equation (\ref{tV}))
for the plateau
formation time as function of the plateau front velocity $V_{\max }$. Several
numerical tests (one of them is shown in Figure 3) have confirmed the
applicability of this model in the solar wind conditions.

In general, our analysis and numerical estimations have revealed that the
quasilinear proton diffusion driven by the observed solar wind turbulence
provides a robust generation mechanism for nonthermal proton tails, which
can explain their routine observations in the solar wind. By implementing
the measured characteristics of the turbulence at proton kinetic scales $\mu
_{1},$ $W_{\mu 1}$, and $p_{2}$ in our model, we estimated the time scale of
the quasilinear plateau formation in the velocity space. For typical
nonthermal tails with $V_{\max }=\left( 1.5-2.5\right) V_{\rm A}$ the
estimated formation time is $0.5 - 3$ h, which is much less than the
solar wind expansion time at 1 AU. This indicates that the local generation
of such tails observed in the solar wind is a natural consequence of the
observed kinetic-scale Alfv\'{e}n turbulence.

Some longer tails may require more time to develop, in which case a
non-local radially-dependent problem has to be solved numerically using
suitable solar wind models. In principle,with sufficient KAW power at $\mu
>\mu _{2}$, the quasilinear plateau can extend well above $3V_{\rm A}$, where an
asymptotic $\mu \gg 1$ analysis by \inlinecite{Rudakov2012} is applicable.
However, the turbulence level is usually insufficient to form the tails with
$V_{\max }>3V_{\rm A}$ within required time scales, which is reflected in the
fast growth of the formation time beyond $t_{\mathrm{SW}}$. This constrain
is compatible with rare observations of such long tails.

The most favorable conditions for the tail generation in terms of $%
V_{{\rm Tp}}/V_{\rm A}$ occur in the range $V_{{\rm Tp}}/V_{\rm A}\simeq 1 - 1.5$ for spectral
slopes ranging from $p_{2}=$2.5 to 4, correspondingly. The optimal $\left(
V_{{\rm Tp}}/V_{\rm A}\right) _{\mathrm{opt}}$ is not much varying with the spectral
slopes in the range of interest, and we can say that in general $\left(
V_{{\rm Tp}}/V_{\rm A}\right) _{\mathrm{opt}}\approx 1$. Such conditions are often met at
1 AU. As is seen from Figure 5, the range of $V_{{\rm Tp}}/V_{\rm A}$ allowing for the
tail generation has two bounds, upper
$\left( V_{{\rm Tp}}/V_{\rm A}\right) _{\mathrm{cr2}}$ and lower
$\left( V_{{\rm Tp}}/V_{\rm A}\right) _{\mathrm{cr1}}$. The upper
bound $\left( V_{{\rm Tp}}/V_{\rm A}\right) _{\mathrm{cr2}}$ is relatively large,
its values for different spectral slopes are very scattered, and it is not
very restrictive for the tail generation. On the contrary, the lower bound
$\left( V_{{\rm Tp}}/V_{\rm A}\right) _{\mathrm{cr1}}$ is more restrictive. It has
more collimated values that are not much different from each other and are not
so small, which is
seen from Figure 5. Say, for $V_{\max }\simeq 1.5V_{\rm A}$ tails $\left(
V_{{\rm Tp}}/V_{\rm A}\right) _{\mathrm{cr1}}\simeq 0.25$ and such tails cannot be
generated locally if $V_{{\rm Tp}}/V_{\rm A}<0.25$. The range of favorable $V_{{\rm Tp}}/V_{\rm A}$ can
not be extended towards smaller values. It seems that only short
tails with $V_{\max }<1.5V_{\rm A}$ can be generated at $V_{{\rm Tp}}/V_{\rm A}\approx 0.1$.
Relatively long tails are expected with steep spectral slopes $p_{2}$
that are positively correlated with turbulence levels.

We also analyzed a particular case reported by \inlinecite{Sahraoui2010}. We
found that the measured turbulence can generate a nonthermal tail extending
to $V_{\max }\lesssim 2V_{\rm A}$, but the proton number density in the tail is
insufficient to explain the measured temperature anisotropy of the protons.
We suggest that the nonlinear resonance broadening and enhanced collisional
diffusion at $V_{z}\lesssim V_{\rm A}$ are important mechanisms significantly
increasing the proton number density in the tail. Yet another possibility
is that the longer tail producing larger measured anisotropy was generated at
shorter radial distances. The related effects require further investigation.

Behind the amplitude-related constrains mentioned above, the tail length and
shape can also be constrained by plasma instabilities. Several instabilities
can develop on the plateau-like proton tail, including the fast magnetosonic,
parallel Alfv\'{e}n-cyclotron \cite{Rudakov2012}, and oblique Alfv\'{e}n
\cite{Daughton1999,Voitenko2003} instabilities.
Although \inlinecite{Daughton1999} and \inlinecite{Voitenko2003} studied
instability driven by the bump-on-tail, its generation mechanism is related
to the anomalous Doppler (proton-cyclotron) resonance rather than the
inverse Landau damping, and hence is not restricted to the bump-on-tail but
can also be efficient with plateau-like distributions. The cyclotron
diffusion of tail protons, driven by these instabilities, can constrain tail
lengths and produce proton-cyclotron quasilinear plateau reported
by \inlinecite{Marsch2001} and \inlinecite{Marsch2011}, or
even produce the bump-on-tail features. In turn, the nonthermal
proton tails and beams can re-emit KAWs
\cite{Daughton1999,Voitenko2003,Wu2012,Nariyuki2012}
and significantly modify the nonlinear evolution of the
high-amplitude circularly-polarised Alfv\'{e}n waves (\opencite{Nariyuki2009},\citeyear{Nariyuki2012}).
A complex interplay of these processes requires further
investigations.

Obtained analytical expressions for the proton diffusion coefficient can be
incorporated in more sophisticated solar wind models accounting for external
forces (electric and gravity), and both the particle-particle (Coulomb) and
the wave-particle (Cherenkov) interaction terms in the right-hand side of
Equation (\ref{Vlasov}). Because of their complexity, such models need to be
solved numerically. Our preliminary simulation results demonstrate that the
kinetic-scale Alfv\'{e}nic turbulence can create such nonthermal tails in
the proton VDFs at the radial distances $\lesssim 20$ solar radii (see
detailed descriptions of these simulations
in our accompanying paper \cite{PiVo2012}).

We did not concern here the bump-on-tail features (proton beams) often
observed in the solar wind. Such distributions have not been reproduced by
the uniform quasilinear diffusion we studied. However, with variable level
of the turbulence launched through the solar wind base, the irregularity of
the generated tail length can easily produce the bump-on-tail features by
the time-of-flight effect. Namely, in a region of enhanced turbulence
the high-velocity end of a dense nonthermal tail propagates faster than the bulk
protons. As a consequence, at larger radial distances this proton population
penetrates a more quiet region with a weaker tail and appears there as a bump on tail.
This mechanism is compatible with Helios observations of time-variable
bumps on virtually persistent tails \cite{Marsch1982}.

In conclusion, appearance of nonthermal proton tails and
beams in the solar wind is unavoidable in the presence of
kinetic-scale Alfv\'{e}nic turbulence with sufficient amplitude.
There are several other generation mechanisms for proton beams and tails (see Introduction),
which can compete with each other or replace in varying solar wind conditions.
More detailed statistical correlation studies are needed to discriminate among them
and to find what mechanism is dominant in the solar wind. It would be particularly
interesting to have a look at correlations between the tail lengths and/or densities
and the turbulence levels at kinetic scales.

\begin{acks}
This research was supported by the Belgian Federal Science Policy Office
(via IAP Programme, project P7/08 CHARM) and
by the European Commission's FP7 Program (projects 263340 SWIFF, and 313038 STORM).
We thank the referee for his constructive criticism that helped us to improve this paper.
\end{acks}


\end{article}

\end{document}